# Non-linear magneto-optic and self polarization rotation by superposition of states

S. Pradhan[1,*], A. Kani[2], H. Wanare[2], R. Behera[1] and A.K. Das[1]

[1]Laser and Plasma Technology Division, Bhabha Atomic Research Center, Mumbai 400 085, India

[2]Department of Physics, Indian Institute of Technology, Kanpur 208 016, India

*Corresponding author email address: spradhan@barc.gov.in

## Abstract

We report the observation of enhanced magneto-optic rotation as the coherent superposition of different hyperfine states is established in an atomic sample. The polarization rotation near the two photon Raman resonance condition appears to have an analogous characteristic to the well established Faraday rotation observed in the vicinity of a single photon resonance, however it contains sharp features arising from coherent population trapping state. The profile of the two photon rotation signal exhibits interesting features for slightly imbalanced circular polarization component of the laser field as well as for on and away from the single photon resonance. The investigation can be used to explore the effect of superposition states generated by coherent population trapping on optical activity. A complete density matrix based numerical simulation that consistently captures all the relevant features of the experiment is presented.

**Introduction:**

The quantum superposition states can be generated by suitably configuring quantum interference effect in the laser-atom interaction process. It has found several uses in fundamental research study like tailoring atomic properties, slow and superluminal light propagation, sub-recoil laser cooling etc [1-4]. It also plays an indispensable role in miniaturized atomic clock and magnetometer [5, 6]. An important aspect of the superposition state generated by coherent population trapping (CPT) is that even though the medium becomes transparent to the light field, yet it remains optically active. Thus, it provides an ideal system to investigate the properties of optical activity arising out of the superposition of states. We bring out the essential differences between polarization rotation arising from one photon transitions in contrast to two-photon CPT transition.

The nonlinear magneto-optic effect (NMOE) observed in laser atom interaction process arise from the bi-polarization nature of the linearly polarized light, where the left and right circularly polarized ($\sigma^-$ and $\sigma^+$) components get differentially phase shifted in presence of non-zero magnetic field. It may be recalled that in NMOE two excitation pathways are simultaneously coupled with a common level, even though CPT is not essential for NMOE it involves similar coupling of two fields to a common excited state. However, in many instances the ground state Zeeman coherence is neglected in formulating the model, which has been successfully used to explain experimental outcome of polarization rotation [7]. The governing mechanism of this model relies on different interaction for the $\sigma^-$ and $\sigma^+$ light interacting with the Zeeman non-degenerate atomic sample. It leads to accumulation of a differential phase shift between the two orthogonal circularly polarized lights propagating through the medium. This is known as circular birefringence and results in the polarization rotation of the linearly polarized light. There are other mechanisms responsible for NMOE and has been extensively discussed by Budker et.al. [8]. The NMOE finds application in magnetometry, parity violation, quantum information processing etc. [8, 9].

It is already demonstrated that the light beam with slightly imbalanced $\sigma^-$ and $\sigma^+$ component interacting with a near resonant atomic medium exhibits polarization rotation even in the absence of magnetic field. This effect, known as polarization self rotation (PSR), arises as the interaction of an imbalanced light with an isotropic medium induces dichroism as well as birefringence simultaneously. In the microscopic picture of PSR, the imbalanced polarization introduces population imbalance among the Zeeman sub-states through differential optical pumping. Thus, the $\sigma^-$ and $\sigma^+$ light propagates in the medium experiencing different refractive index resulting in a net phase shift among them. The consequent polarization rotation gives rise to PSR, which becomes prominent for resonant optical pumping saturation parameter $\kappa \ll 1$ for the atomic vapor in the collision free regime [10]. The differential light shift arising due to the imbalanced circular polarization components also contributes to the PSR, even for atomic structure where optical pumping is precluded. The stark shift based PSR has a prominent role for higher light intensity. Recently, PSR is studied in ultra-cold Rubidium atoms in presence of a small static magnetic field [11]. It is worth noting that NMOE requires a non-zero magnetic field unlike PSR which occurs irrespective of the magnetic field.

In many of these studies, NMOE and PSR have been explained by assuming Zeeman coherence and hence the superposition between Zeeman sub-states. In contrast to these studies near single photon resonance, we have investigated polarization rotation of the bi-chromatic field near the two-photon Raman resonance condition also known as dark resonance or coherent population trapping (CPT) state. One of the advantages of using dark resonances emerges from the narrower (sub-natural linewidth) resonance width as compared to the Zeeman splitting width even for small magnetic field of ~100 mG. In this work we have applied a magnetic field of ~600 mG. This allows us to distinguish the polarization rotation occurring due to various assemblies of states having distinct Raman resonance in a Zeeman non-degenerate system. The relevant energy level diagram of the $^{85}$Rb interacting with a bi-chromatic linearly polarized light in presence of a magnetic field is illustrated in the Figure-1(a). The various CPT mechanisms that are simultaneously responsible for the polarization rotation are identified with a simplified energy level scheme depicted in Figure-1(b). Here, the linearly polarized light is split into equal $\sigma^-$ and $\sigma^+$ component. It is apparent that even in this simplified scheme, the dynamics is complex as competing CPT states occur at the same two photon Raman resonance condition. However one can broadly categorize the polarization rotation contribution from two groups. For the first kind, both $\sigma^-$ and $\sigma^+$ components drive either one or two independent CPT states. In the second category, only one of the $\sigma^-$ and $\sigma^+$ component is part of a CPT family. Both the situations give rise to polarization rotation preserving the sub-natural line width of the dark resonance. The overall dynamics is complicated due to an interplay of several mechanisms like Zeeman & hyperfine optical pumping, light shift, Doppler shift due to thermal motion, Zeeman shift of sub-state, differential Clebsch-Gorden coefficient, single-photon detuning etc.. In this work, further complexity of single photon processes to the polarization rotation is completely eliminated by FM spectroscopy and the sole contribution of two-photon process is measured.

The CPT state plays a critical role in obtaining the large rotation in the proximity of the two-photon resonance which can be individually resolved at the five different frequencies in presence of the magnetic field. It is the subnatural linewidth associated with the CPT dip that allows such spectral resolution. Away from the two-photon resonance single photon effects dominate. For the single photon process, broad spontaneous emission lineshape (~ 6 MHz) completely subsumes within it the different resonant responses arising due to the various Zeeman shifted levels in presence of the above mentioned magnetic field. One obtains smaller magnitude of polarization rotation in comparison to the two-photon resonance case. In order to distinguish the effects arising due to the various Zeeman levels one would require rather strong magnetic field whose Larmour frequency is larger than the combined decays of the upper and lower levels, and the inhomogeneous Doppler broadening for the atomic gas at room temperature.

The situation at the two-photon resonance is significantly different. One obtains five distinct frequencies where one obtains two-photon resonance in presence of the applied magnetic field. These distinct frequencies occur for a combination of the ground state $m_F$ values corresponding to the $F = 2$ and $F = 3$ ground state manifold. The NMOE effect is significant at the two-photon Raman condition and thus the ground state energy difference determines the resonance frequency whereas the excited state $m_F$ plays little role. Multiple combinations of set of CPT resonances arising out of the two fields and three atomic states contribute simultaneously to produce the NMOE. The two-photon NMOE signal is obtained at five distinct frequencies due to multiple CPT resonances which occur at five values of the ground state combination ( $F = 3, m_F$ and $F = 2, m_F$ ) with $m_{F, F=3} + m_{F, F=2}$ = -4,-2,0,+2,+4. For the case of $m_{F, F=3} + m_{F, F=2}$ = -4 and +4 three CPT configurations contribute to NMOE whereas for $m_{F, F=3} + m_{F, F=2}$ = -2, 0 and +2 four CPT configurations contribute. The four different types of CPT for specific ground state $m_F$ values are indicated in the Fig. 1(b) in different colors (online version). The NMOE signal results from such simultaneous contributions from these competing CPT processes.

We present results of the theoretical model that captures the dynamics of this coupled 19 level atomic system using the semiclassical density matrix formulation. The theoretical model clearly captures all the essential features of the experiment. We would like to point out that at resonance (both single and two-photon resonance) we obtain a dispersive response with zero rotation at the line center of the CPT dip, however even a small single photon detuning (~ 0.5 of the natural line width) leads to the peak/dip like response which matches with the experimental results. The appropriate cross terms of the density matrix are calculated in the steady state to obtain the refractive index and its difference allows one to calculate the polarization rotation. The degree of thermal redistribution of the ground state population has been included in the theoretical model to match the experimental measurement, however Doppler broadening has not been included in the present theoretical model. The details of the theoretical treatment and its extensions discussing all the associated issues related to the two-photon Raman resonance, the effect of Zeeman coherence, collisional decay of the ground state population and coherences, the effect of Doppler broadening etc. will be discussed in detail elsewhere. These experimental and theoretical studies will be helpful in exploring the dynamics resulting from the dark superposition state.

**Experimental system:**

The schematic of the experimental set-up is described in the Figure-2. The experiment is carried out with a surface emitting DBR diode laser at 780 nm. The output laser beam is linearly polarized, emitting 978 µW and has an isotropic Gaussian width of ~1.5 mm. A part of the laser beam (~10%) is used for the purpose of frequency stabilization. The rest of the laser beam is expanded three times to a Gaussian width of ~4.5 mm. The expanded beam is passed through a half wave plate and a polarizing beam splitter cube for controlling the transmitted light intensity and maintaining the polarization purity (linear polarization). We put an additional (as per requirement) retardation plate before sending the beam through the Rubidium vapor cell and finally detect the orthogonal linear polarization component of the transmitted beam by a pair of silicon photo-detector. The polarization beam splitter cubes used as polarizer and analyzer in this experiment are specified for transmitted horizontal polarization >92% and reflected vertical polarization >99%. The atomic cell contains Rubidium in natural isotopic composition and is free from buffer gas. The Rubidium cell is enclosed inside a magnetic shield that is used to nullify the effects of any external magnetic field and a solenoid is used to apply a controlled magnetic field. The atomic cell temperature is maintained at $35^0$ C. The bi-chromatic field is generated by modulating the laser frequency at half of the ground state separation (~1.5178 GHz) of $^{85}$Rb with amplitude of -1.8 dBm using a radio frequency generator. The laser frequency is locked to the middle of $5s_{1/2}F = 3 \rightarrow 5p_{3/2}F'$ and $5s_{1/2}F = 2 \rightarrow 5p_{3/2}F'$ transition of $^{85}$Rb atom using wavelength modulation spectroscopy. The frequency of the RF generator is modulated with a sinusoidal wave of 900 Hz with amplitude of 900 mV. The transmitted and the reflected component of the bi-chromatic laser light are phase sensitively detected with respect to the frequency modulation applied to the radio frequency. The RF oscillator output frequency is digitally scanned across the Raman resonance condition by a computer equipped with a data acquisition and control card.

**Results and Discussions:**

The transmitted and the reflected component of the signal are phase sensitively detected with a pair photo-detectors as a function of the two photon (Raman) detuning without the wave plate before the Rubidium cell (please see Fig.-2). The experimental geometry is identical to the NMOE experiments, where the probing laser field is linearly polarized, and is made to interact with the medium and finally the polarization rotation is investigated by an analyzer; except a bi-chromatic field is used in place of the conventional monochromatic probe field. It may be noted that there will be polarization rotation at both the frequency components of the bi-chromatic field. These single photon contributions to the signal are filtered out by phase sensitive detection with respect to the applied frequency modulation to the radio frequency. Thus the resultant signal has contribution arising purely out of the two photon process to the polarization rotation and is investigated as a function of the detuning of the Raman resonance condition. The physical multi-photon interaction mechanism of the process involved is shown in the schematic diagram illustrated in Fig.1. In the simplified level scheme shown in Fig.1(b), the $\sigma^-$ ($\sigma^+$) pair of light field shown by the dotted (solid) line constitute various three level "Λ" systems simultaneously. As the two photon Raman resonance condition is satisfied, all the interacting atoms are pumped into their respective non-absorbing dark states [12]. The real part of the first order susceptibility $\chi^1$ responsible for the refractive index depends on the Rabi frequency of the respective excitation. The physical mechanism of differential refractive index for $\sigma^+$ and $\sigma^-$ giving rise to polarization rotation for two photon processes is analogous to Faraday rotation and Hanle effect for single photon process discussed by Budker et.al. [8], however it involves bichromatic fields creating and probing various ground state coherent superpositions. The polarization rotation is realized in such a coherently prepared atomic sample, thereby offering the possibility of exploring intriguing aspects of atomic coherence and taking advantage of the associated ultra-narrow linewidth accompanied by the enhanced non-linear susceptibility.

The polarization rotation signal illustrated in Figure-3(a) has the transmitted signal ~1000 times stronger than the reflected signal. The transmitted signal essentially provides a measure of the overall CPT signal, whereas the reflected signal is the NMOE signal. The five dispersive CPT signals shown in Fig-3(a) are generated as the Raman resonance is achieved at five different modulation frequencies as discussed earlier. It should be noted that corresponding to each CPT condition the corresponding NMOE signal is generated. Thus similar to CPT, the NMOE signal is exclusively generated by establishment of the superposition of the different hyperfine ground states. The signal to noise ratio (SNR) of NMOE is improved compared to the CPT signal even though the former has 3 orders of magnitude weaker signal. The improved SNR of the NMOE signal is a consequence of the signal recovery from nearly zero background, whereas the CPT signal is extracted from a relatively large background. The better SNR of the NMOE can be gainfully utilized for the improving sensitivity of magnetometers as compared to that using direct CPT signal. As the CPT and NMOE signals are generated using frequency modulation spectroscopy [13], they represent derivative of the corresponding signal profile. The actual line shape is reconstructed following numerical integration [14] as shown in Figure-3(b).

The NMOE signal has comparable width to the CPT signal. The observation of opposite polarity of the NMOE for $m_F$ = -ve and $m_F$ = +ve pairs of states is an outcome of the advancement and delay of the $\sigma^-$ light compared to the $\sigma^+$ light, respectively, and is a confirmation of Faraday rotation by the superposition of different hyperfine states. The properties of the NMOE have been extensively studied near single photon resonance. Qualitatively extending the observations to two photon resonance condition, one would expect single peak profile for $m_F \neq 0$ sets of ground state and zero contribution for $m_F = 0$. Instead, our experimental result substantially deviates from it as illustrated in Fig. 3(b). The possible distortion of the signal can arise from imbalanced $\sigma^-$ and $\sigma^+$ component, presence of residual vertical polarized light after the polarizer, finite angle between the polarizer and the analyzer, and non-zero reflection of the horizontally polarized light by the analyzer. Such imperfections also result in addition of a small CPT component in the NMOE signal. These defects can be corrected by fine optical alignment and using additional retardation plates, however one is limited by the efficiency of the analyzer.

The imbalance in the circular polarization components is corrected by placing a quarter wave plate before the atomic cell. The profile of the reflected signal is found to be strongly dependent on the angle of the quarter wave plate. The angle of the quarter wave plate is adjusted with a resolution of $0.01^0$ and a nearly pure dispersive profile of NMOE is obtained as shown in curve (a) of Figure-4. It is worth noting that the dispersive profile for $m_F$ = -ve and $m_F$ = 0 have the same slope as the CPT and the $m_F$ = +ve have the opposite slope. This structure is also clearly captured by the theoretical model. As the applied magnetic field direction is rotated by $180^0$, the slope of all NMOE profiles change sign except for the $m_F$ = 0 component (This result is not shown). The use of quarter wave plate allows a controlled improvement of the balance between $\sigma^-$ and $\sigma^+$ components of light. Using the theoretical calculation we infer that optical pumping significantly contributes to the relative signal amplitude of the polarization rotation signal obtained at various Raman resonances.

The imbalanced $\sigma^-$ and $\sigma^+$ component polarization also induces PSR, where polarization ellipse of light undergoes self rotation resulting in the change of the reflected signal from the analyzer. The possible operation of PSR for coupling scheme shown in Fig-1 is an interesting problem. In view of the suggestion in Ref. [10] for single photon resonance, our experimental parameters are suitable for observation of PSR as the buffer gas free cell ensures the atomic medium to be in near collision less regime. The reflected signal profile is found to dramatically change as the laser polarization is changed to elliptical polarization by changing the quarter wave plate by ~ $\pm 1^0$ as shown in Figure-4. The observation of peak profile indicates the actual profile (modified due to signal recovery by FM modulation technique) to be dispersive in nature. This is consistent with the prediction of PSR near single photon resonance [10]. The mechanism for such polarization rotation is governed by the imbalanced ground state population induced by elliptically polarized laser assisted optical pumping. The other PSR mechanism discussed for the single photon process rely on differential AC stark shift of the atomic state due to imbalanced light intensity of the $\sigma^-$ and $\sigma^+$ components. However, theoretical investigations will be presented elsewhere which reveal the exact interplay of various mechanisms behind the observed PSR/NMOE kind of optical rotation near the two-photon resonance.

The comparison of NMOE and PSR kind of observation near two photon resonance with the single photon process is subtle. However similar to single photon resonance, optical pumping leads to sharing of unequal population among the simplified asymmetric set of few-level models illustrated in Fig.1(b). Though, these are accompanied by a series of simultaneous, coupled lambda systems. These imbalanced ground state populations can introduce differential refractive index between $\sigma^-$ and $\sigma^+$ components. It may be noted that the different single photon resonance frequency of $\sigma^-$ and $\sigma^+$ beam in Zeeman non-degenerate system is no longer valid as the two photon resonance is simultaneously satisfied for both $\sigma^-$ and $\sigma^+$ beams. Despite these differences, the similarity in the observed polarization rotation near single photon and two photon resonances is interesting.

The results of the theoretical model based on density matrix formulation presented in Figure-5 is consistent with the experimental results shown in Fig.-4. It may be noted that the Faraday rotation (PSR) signal due to CPT state has a dispersive (peak/dip) line shape as a function of two photon resonance for zero single photon detuning. However as one moves away from the single photon resonance by about $\Gamma/2$ ($\Gamma$ =natural line width), these dispersive (peak/dip) line shapes of the Faraday rotation (PSR) signal transform into peaks or dips (dispersive), these dips/peaks (dispersive) continue to occur even for larger single photon detuning. This is consistent with the experimentally obtained signal profile, as FM spectroscopy was implemented for recovery of purely the two-photon signal. The NMOE signal is significant only at the two-photon resonant condition thus highlighting the important role played by the superposed ground states at the CPT condition. To match with the experimental results we have used single-photon detuning of $-\Gamma$ corresponding to a frequency shift of ~ 6 MHz.

The relative strengths of the NMOE resonances is not only dependent on the various Clebsch-Gorden coefficients along the respective transitions but also on the remnant optical pumping that continues to exists in presence of the thermal and collisional redistribution of the population between the ground states. In order to overcome the dominant optical pumping effects we introduce phenomenological incoherent decay rates so as to obtain thermal distribution of population corresponding to room temperature Boltzmann distribution in absence of the coherent optical fields. Introduction of these incoherent decay processes between the various states results in line shape broadenings of the order of $\Gamma$.

Furthermore our theoretical model does not point towards any simplified few-level model [Fig. 1(b)] that captures the main experimental results. We believe that the excellent match of the theoretical results with the experimental data has been obtained because of inclusion of the couplings of all the 19 energy states arising from the incoherent relaxation processes as well as the coherent optical field couplings.

**Conclusions:**

Polarization rotation by a linearly polarized bi-chromatic light field near two photon resonance is investigated. The contribution of two photon process to the polarization rotation is isolated from the single photon process by using FM spectroscopy. The narrow resonance line width (~30 KHz) allows us to investigate the effects of coherent superposition of the ground states to NMOE. The differential refractive index of the superposition of hyperfine states involving $\sigma^-$ and $\sigma^+$ light is described as the physical mechanism behind the observed NMOE for purely linearly polarized light at two photon resonance. The polarization rotation arising from both the NMOE and PSR at the two-photon resonance is reasonably well captured by the theoretical treatment using density matrix formulation for the 19 level atomic states coupled to the bi-chromatic field. The effects of various family of CPT states acting together is clearly identified. The investigation can play important role for parity violation experiment, atomic clock, magnetometer and many other research efforts that could explore the effect of such superposition of states.

**Acknowledgement:**

The authors are thankful to Dr. L.M. Gantayet for discussion and support during this research work. The fruitful suggestions of the referee are thankfully acknowledged.

**References:**

[1] A.V. Taichenachev, V. I. Yudin, V. L. Velichansky, A.S. Zibrov, and S.A. Zibrov, Phys. Rev. A, **73**, 013812 (2006).

[2] L. V. Hau, S.E. Harris, Z. Dutton and C.H. Behroozi, Nature, **397**, 594-598 (1999).

[3] L.J. Wang, A. Kuzmich and A. Dogariu, Nature, **406**, 277-279 (2000).

[4] A. Aspect, E. Arimondo, R. Kaiser, N. Vansteenkiste, C. Cohen-Tannoudji, Phys. Rev. Lett., **61**, 826-829 (1988)

[5] S. Knappe, V. Shah, P.D.D. Schwindt, L. Hollberg, J. Kitching, L Liew and J. Moreland, App. Phys. Lett., **85**, 1460-1462 (2004).

[6] P.D.D. Schwindt, S. Knappe, V. Shah, L. Hollberg, and J. Kitching, Appl. Phys. Lett., **85**, 6409 (2004).

[7] G. Labeyrie, C. Miniatura, and R. Kaiser, Phys. Rev. A, **64**, 033402 (2001).

[8] D. Budker, W. Gawlik, D.F. Kimball, S.M. Rochester, V.V. Yashchuk, and A. Weis, Rev. Mod. Phys., **74**, 1153 (2002)

[9] M. Atature, J. Dreiser, A. Badolato, and A. Imamoglu, Nature Physics, **3**, 101 (2007).

[10] S.M. Rochester, D.S. Hsiung, D. Budker, R.Y. Chiao, D.F. Kimball, and V.V. Yashchuk, Phys. Rev.A, **63**, 043814 (2001).

[11] T. Horrom, S. Balik, A. Lezama, M .D. Havey, and E.E. Mikhailov, Phys. Rev. A, **83**, 053850 (2011).

[12] M. Fleischhauer, A. Imamoglu, and J. P. Marangos, Rev. Mod. Phys. **77**, 633 (2005).

[13] G.C. Bjorklund, M.D. Levenson, W. Lenth, and C. Ortiz, Applied Phys. B, **32**, 145 (1983).

[14] H. Li, V.A. Sautenkov, T.S. Varzhapetyan, Y.V. Rostovtsev, and M.O. Scully, J. Opt. Soc. Am. B, **25**, 1702-1707 (2008)

**Figure Captions:**

**Figure-1: (a)** The schematic of the coupling scheme for $^{85}$Rb D-2 transition excited by a bi-chromatic linearly polarized light in presence of a static magnetic field. The linearly polarized light is represented as a combination of right and left circularly polarized light. There can be five Raman resonances in a Zeeman non-degenerate system. **(b)** The simplified Energy level diagram at one of the Raman resonance. The gray horizontal line corresponds to the virtual level. The solid (dashed) arrow corresponds to $\sigma^-$ ($\sigma^+$) component of the light field. The several competing CPT configuration contributing to the polarization rotation are shown by different color (online only).

**Figure-2:** Schematics of the experimental set-up for study of NMOE and PSR under CPT. The surface emitting DBR laser shown as SE-DBR is frequency modulated with a radio frequency generator to produce a bi-chromatic field. A part of the laser light is used to stabilize the laser frequency at desired position near the atomic resonance. The modulation frequency is tuned to generate quantum superposition state in $^{85}$Rb atom. The transmission light intensity and polarization are monitored using an analyzer. The signal to noise ratio are improved by frequency modulation and phase sensitive detection technique.

**Figure-3(a):** The transmitted (gray line) and the reflected (black line) signal as a function of frequency difference between the bi-chromatic fields and their relative amplitude are shown by the right and left ordinate respectively. The shown abscissa has an offset of 1.517 GHz. The transmitted part of the signal is identical to the CPT signal and the reflected fraction is resulted from polarization rotation.

**Figure-3(b):** The integrated profile of transmitted (dashed gray line) and reflected (solid black line) fraction of the laser light as shown in the Fig.-3(a). The signal to noise ratio looks improved as compared to Fig-3(a) data as the noise is averaged out during numerical integration. Since the data are taken using wavelength modulation technique, the integrated data represent the actual signal profile.

**Figure-4:** Dependence of the reflection signal on the degree of elliptical polarization of the light field as a function of Raman resonance condition for various assemblies of superposition of different hyperfine states. All three plots are in the same scale but their offset in the ordinate axis is shifted for better visualization. (a) The residual imbalance in the cross-circular polarization component is removed by critically aligning the axis of a quarter wave plate placed before the cell. (b) Polarization imbalance ($\sigma^+ > \sigma^-$) is imposed by rotating the quarter wave plate by $\sim +1^0$. The observed profile is a convolution of PSR, NMOE, and CPT signal but dominated PSR. (c) Same as (b) but $\sigma^-$ component is larger than $\sigma^+$ component obtained by rotating the quarter wave plate by $\sim -1^0$.

**Figure-5:** Theoretical result of polarization rotation near two photon resonance using semi classical density matrix analysis. The model considered all 19 Zeeman sub-states involved in $5s_{1/2}F = 3 \rightarrow 5p_{3/2}F' = 3$ and $5s_{1/2}F = 2 \rightarrow 5p_{3/2}F' = 3$ transition. The illustrated polarization rotation signal corresponds to the derivative of the actual signal profile and is consistent with the experimental results shown in Fig-4. The curves (a), (b) and (c) corresponds to $\sigma^+ = \sigma^-$, $\sigma^+ > \sigma^-$ and $\sigma^+ < \sigma^-$ respectively, where the imbalance corresponds to $\pm 1^0$ rotation of the quarter wave plate. The vertical lines correspond to the position of various Raman resonances.

**Figures:**

**Figure-1:**

**Figure 1(a)**:

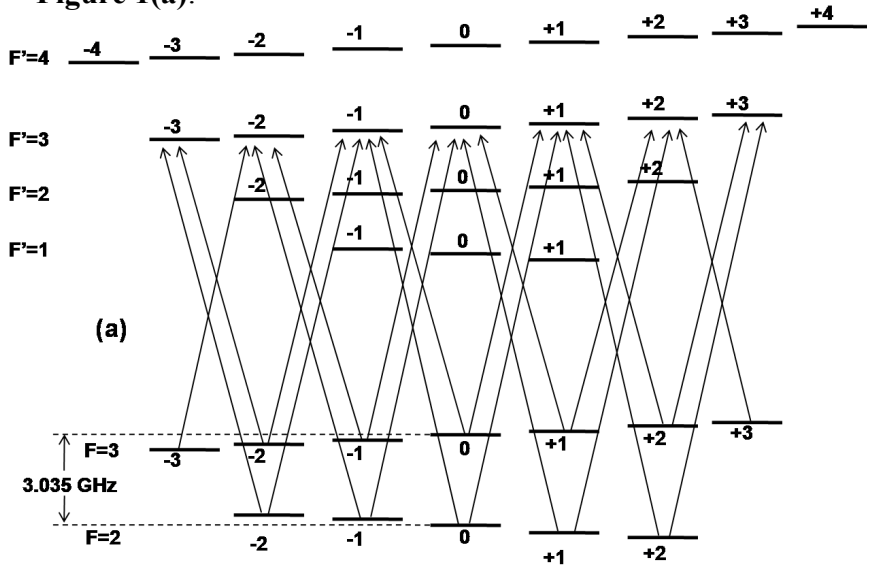

**Figure1(b):**

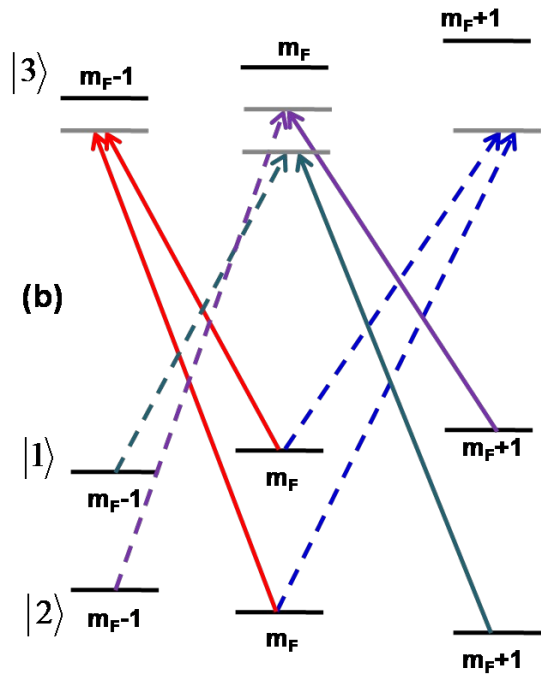

**Figure-2:**

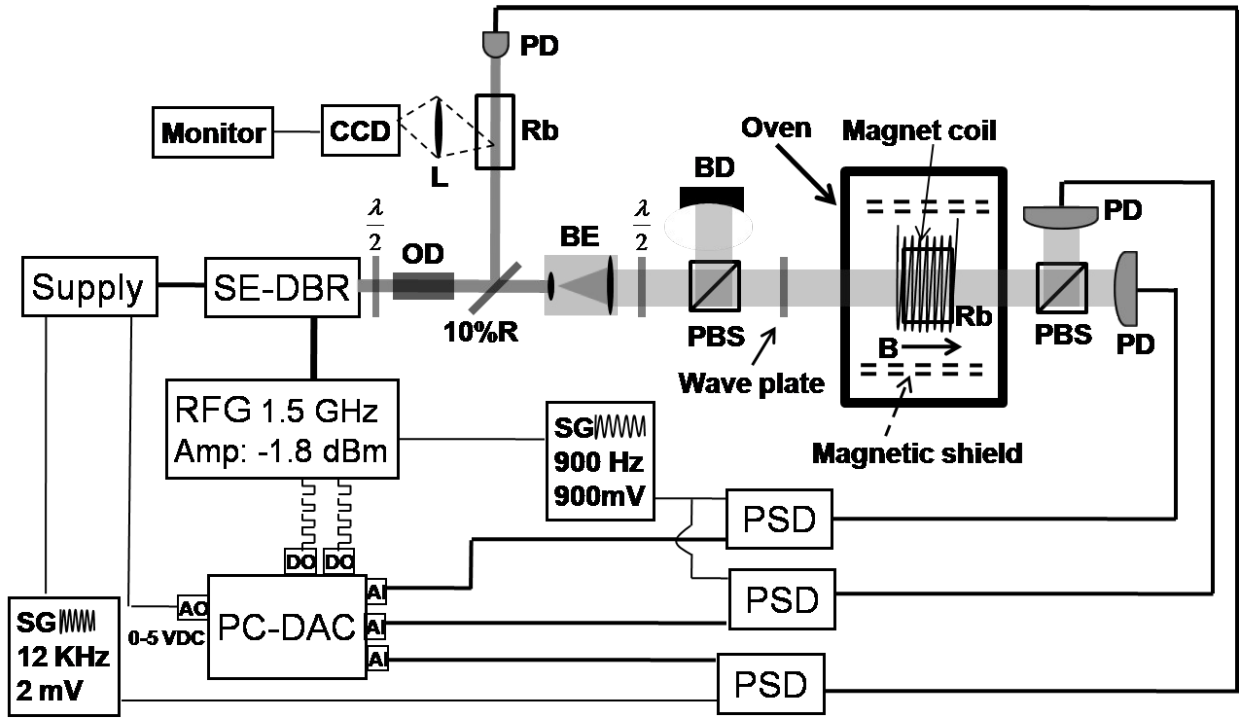

**Figure-3(a):**

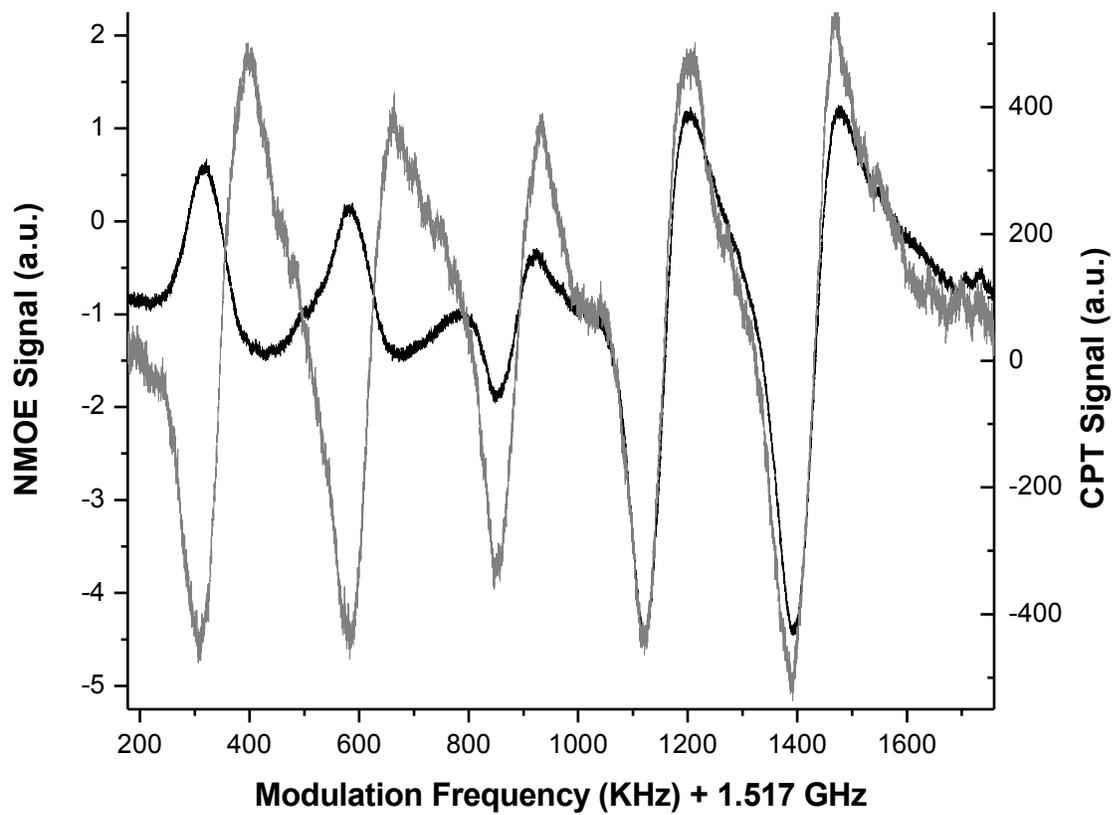

**Figure-3(b):**

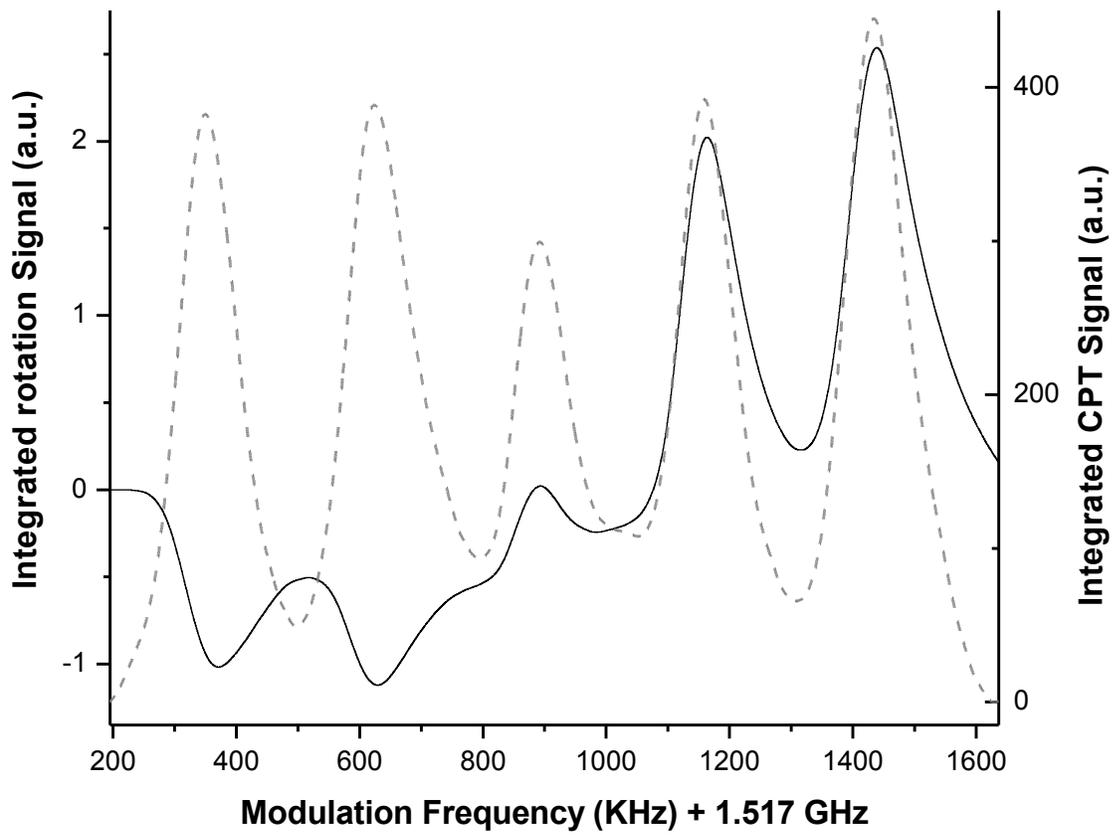

**Figure-4:**

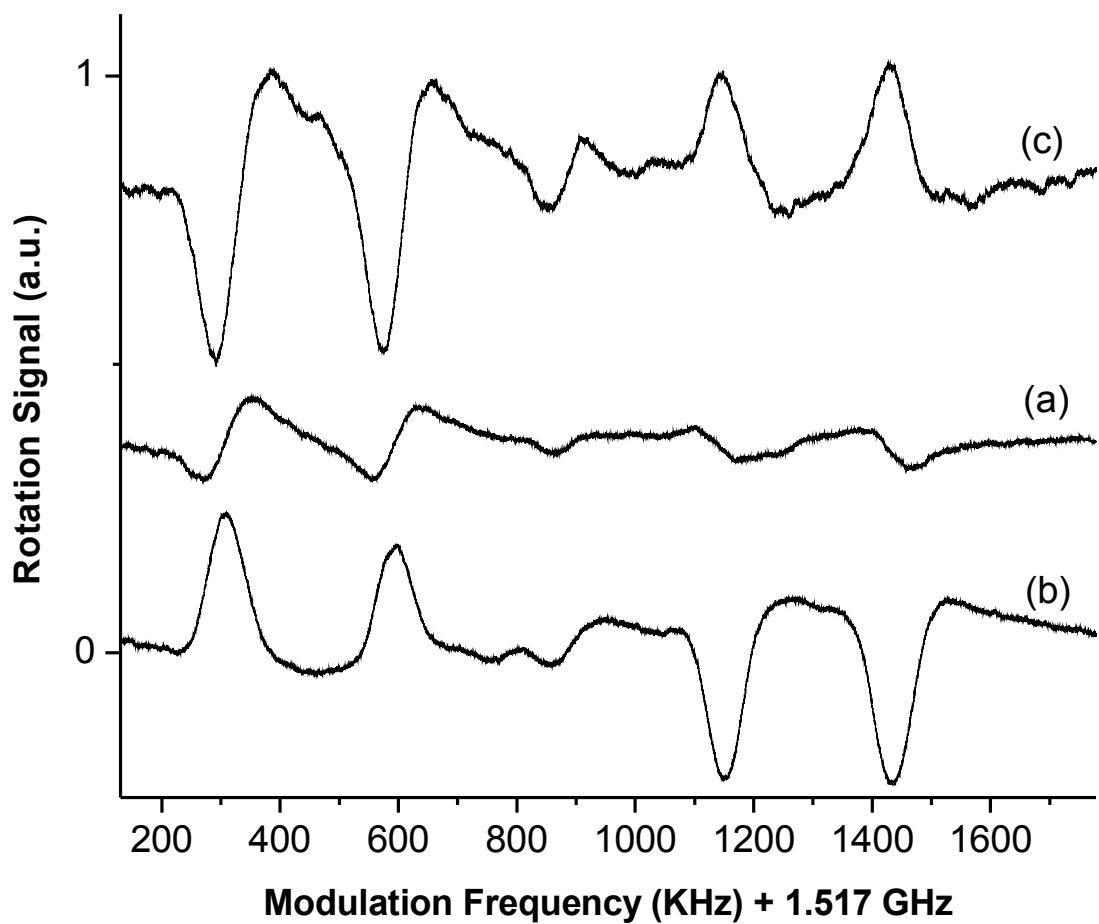

**Figure-5:**

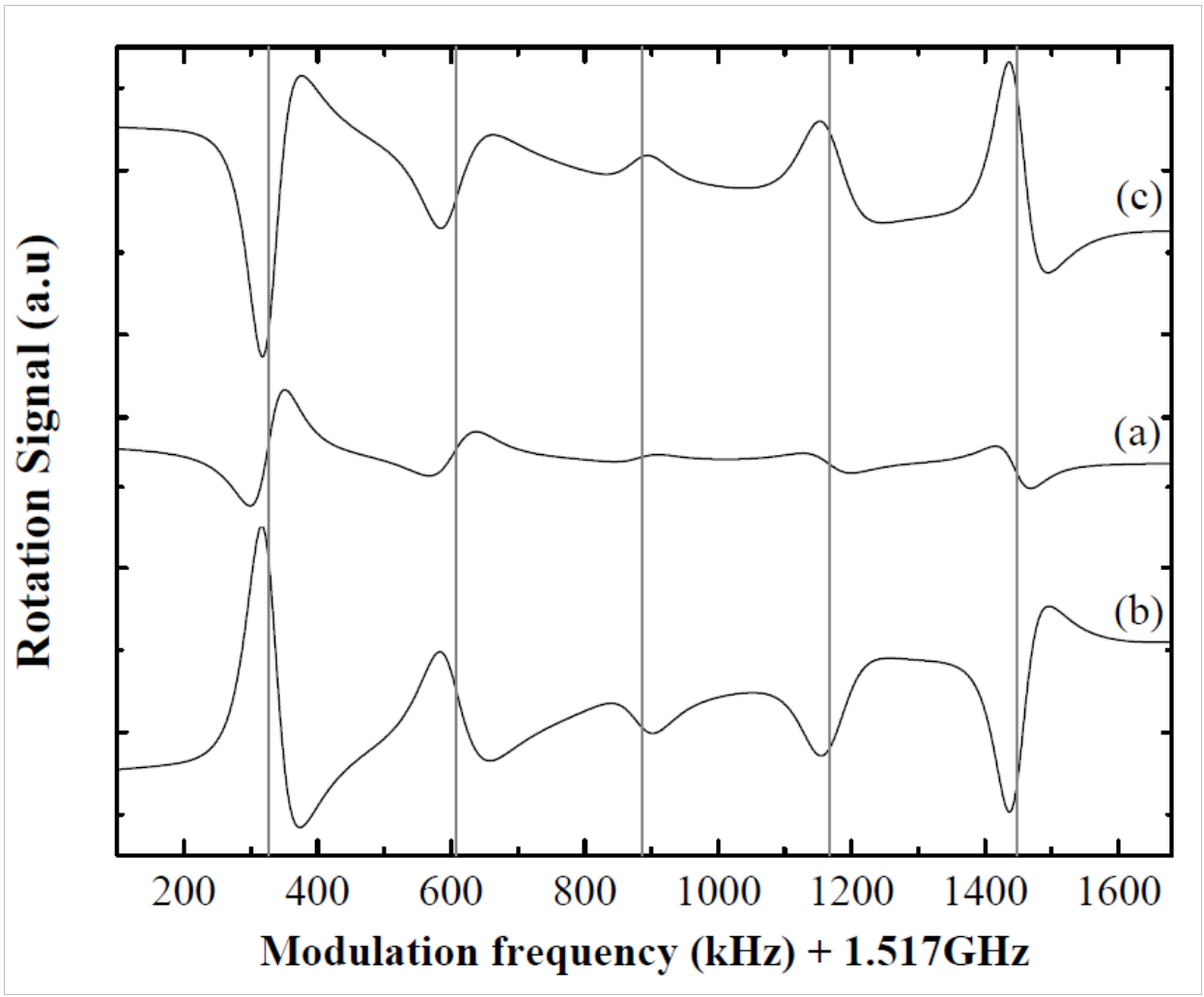